\documentclass[rsi,aip,reprint,amsmath,amssymb]{revtex4-2}
\usepackage{graphicx}
\usepackage{listings}
\usepackage{xcolor}
\usepackage{hyperref}
\usepackage{booktabs}
\usepackage{tikz}
\usetikzlibrary{arrows.meta,calc,positioning}
\tikzset{
  tnode/.style={circle, draw, inner sep=1pt, minimum size=4.2mm, font=\scriptsize},
  tedge/.style={-{Latex[length=1.4mm]}, shorten >=0.5pt},
  dedge/.style={-{Latex[length=1.4mm]}, dashed, shorten >=0.5pt, font=\tiny},
}

\begin{document}

\title{Formally Verified Lock-Free Software Transactional Memory for Scientific Measurement}

\author{Kentaro Kitagawa}
\email{kitag@issp.u-tokyo.ac.jp}
\affiliation{The Institute for Solid State Physics, University of Tokyo, Kashiwa, Chiba 277-8581, Japan}

\begin{abstract}
Automated measurement of condensed-matter experiments requires instrument-control,
data-acquisition, and user-interface threads to access shared, naturally hierarchical state concurrently.
Coarse-grained locking can delay acquisition and cause sample loss, whereas fine-grained locking
 requires deadlock-prone lock ordering across instruments.
Instead, we describe the lock-free software transactional memory (STM) that has been at the
 core of an open-source measurement platform for 16 years, in nuclear magnetic
 resonance experiments and, more recently, in optically detected magnetic resonance experiments.
The STM organizes this state as a tree and provides atomic subtree updates and consistent subtree snapshots.
After initial bundling, an unchanged subtree snapshot is acquired in $O(1)$ time
 through a custom lock-free atomic shared pointer.
Within each bounded TLA+ configuration, TLC exhaustively checks the state space,
 establishing the safety and livelock-freedom properties specified for that configuration.
 Bounded executions of the atomic shared-pointer implementation are separately
 checked under the C11 weak-memory model.
The same Snapshot and Transaction interfaces are exposed to Python scripting and AI-assisted automation.
\end{abstract}

\maketitle

\section{Introduction}
\label{sec:intro}

\begin{figure}
\includegraphics[width=\columnwidth,trim=14 160 7 25,clip]{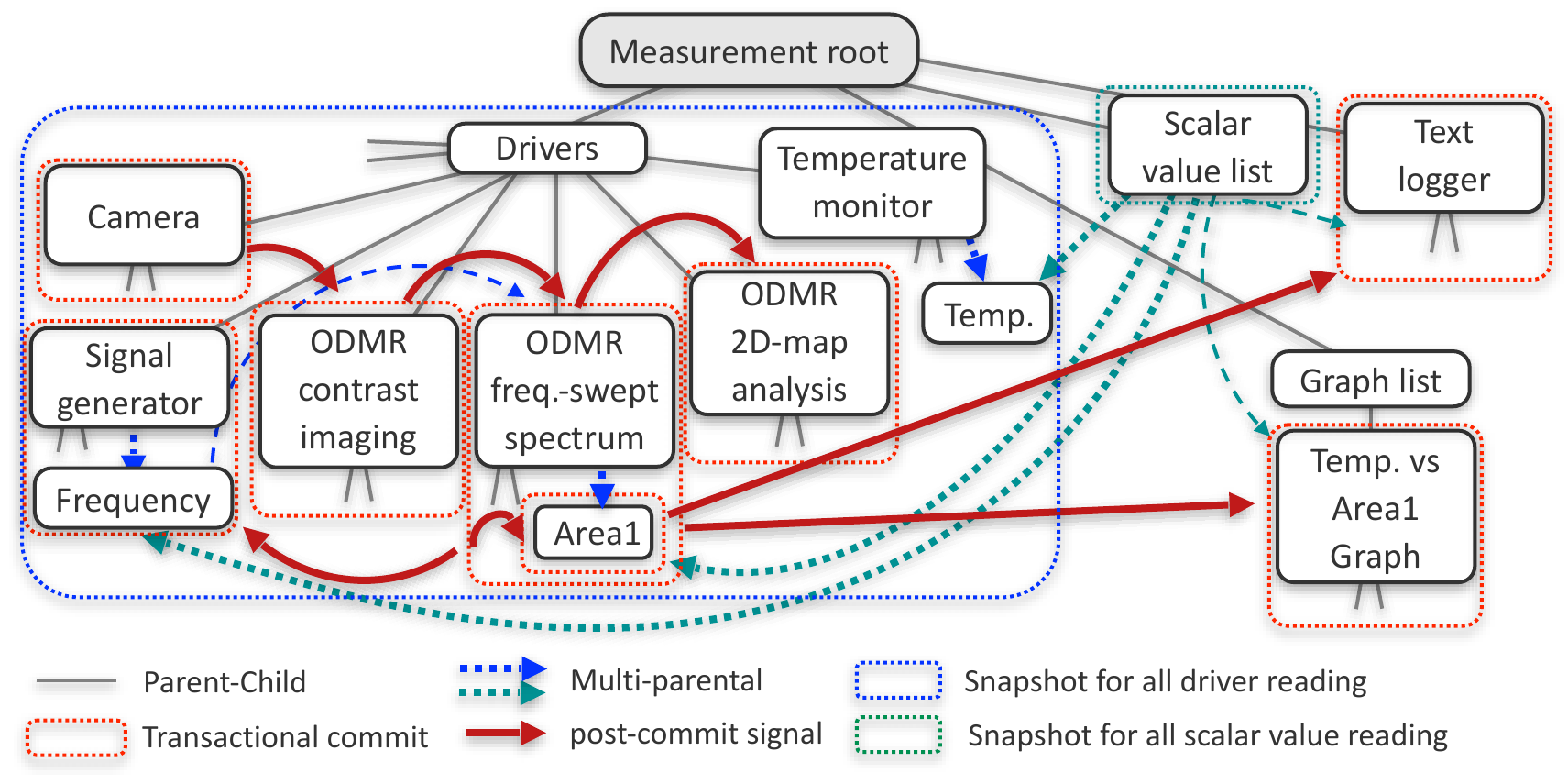}
\caption{Concurrency model of the wide-field optically-detected magnetic resonance (ODMR) measurement of
Fig.~\ref{fig:screenshot}, drawn as its simplified node tree (see legend).
Grey \emph{Parent--Child} edges form the ownership tree of instrument
drivers and display/aggregation sections under the root
\texttt{Measurement} node. The scalar entries (\texttt{Area1},
\texttt{Temp.}, \texttt{Frequency}) are \emph{multi-parented}: each is
owned by its source driver and simultaneously aliased into the Scalar
value list. Each dotted red box is one transactional
commit, atomic over its subtree. When a driver commits, a
\emph{post-commit signal} propagates along the dependency chain; each
listener takes a non-blocking consistent snapshot of its inputs (two
snapshot scopes shown) and atomically commits its own result; each stage
is thus individually consistent, with no lock or transaction spanning the
whole cascade.}
\label{fig:nodetree}
\end{figure}

Measurement software for automated condensed-matter experiments maintains large shared states: measurement parameters, 
acquired data, and instrument conditions. These states are naturally hierarchical.
Instruments own channels and parameters, and analysis routines jointly read the states of multiple instruments.
KAME organizes them as a \emph{tree-structured shared model} (Fig.~\ref{fig:nodetree}).
Many threads operate on this tree concurrently: instrument drivers polled asynchronously 
at intervals on a typical order of 10~ms, higher-rate real-time AD/DA threads, GUI updates, and user scripts.
Collectively, these components require two operations at low latency,
a {consistent snapshot of a subtree} and an {atomic update of a related set of nodes}.
Figure~\ref{fig:nodetree} illustrates both operations on the actual node tree of an ODMR (optically detected magnetic resonance) measurement.
An analysis step reads a camera frame and its metadata together with the source frequency associated with that frame.
The subsequent sweep step atomically advances the source frequency and appends the corresponding spectrum point,
while the GUI concurrently displays the camera image and the frequency-swept spectrum.

A straightforward implementation based on coarse-grained mutual exclusion exposes problems
that are common in experimental practice.
A coarse lock on the shared tree delays acquisition processing behind GUI or scripting activity and,
 once acquisition buffering is exhausted, loses samples.
Finer-grained per-instrument locks reduce such unrelated blocking,
 but introduce lock-ordering constraints whenever an operation spans multiple instruments,
and the ordering discipline must be maintained as new instruments and cross-instrument operations are added.
A further trade-off concerns long-lived consistent views: 
holding a lock for the lifetime of a view blocks writers until the reader finishes,
potentially indefinitely if the reader is an unbounded user script.
Copying the view while holding the lock and then releasing it avoids such blocking, 
but requires the subtree to be traversed and copied for every snapshot.

Software transactional memory (STM) offers a structural solution.
Accesses to shared data are grouped into transactions whose committed updates appear atomic to other threads;
a conflicting transaction is aborted and retried by the runtime or the transactional API, 
without exposing its intermediate states.
Lock-order deadlock is eliminated by design and there is no need to manage lock ordering,
although STM by itself does not eliminate livelock or starvation.

The motivation is not raw throughput. 
What measurement software needs from an STM is the two operations above---consistent subtree snapshots that hold no locks on the live state,
 and subtree-atomic updates---together with sustained progress under contention.
Existing STM systems offer differing progress guarantees and do not by themselves ensure this combination.\cite{shavit1997stm,herlihy2003dstm,harris2005composable,dice2006tl2,hickey2008clojure}
KAME provides this combination: a reader retains an immutable subtree snapshot without holding any lock, 
and an unchanged subtree is reused from its previously bundled view rather than recopied for every read.

In this paper, we report an open-source measurement framework for condensed matter experiments built on a tree-structured variant of STM,\cite{kame}
together with the formal verification of its lock-free implementation.
KAME's STM (2010--) written in C++ is distinguished by its unique features of tree-structured organization and subtree consistency
 via the \emph{bundle protocol}, integrated with signaling of a subtree snapshot on transactional commit.

\begin{figure*}
\includegraphics[width=\textwidth]{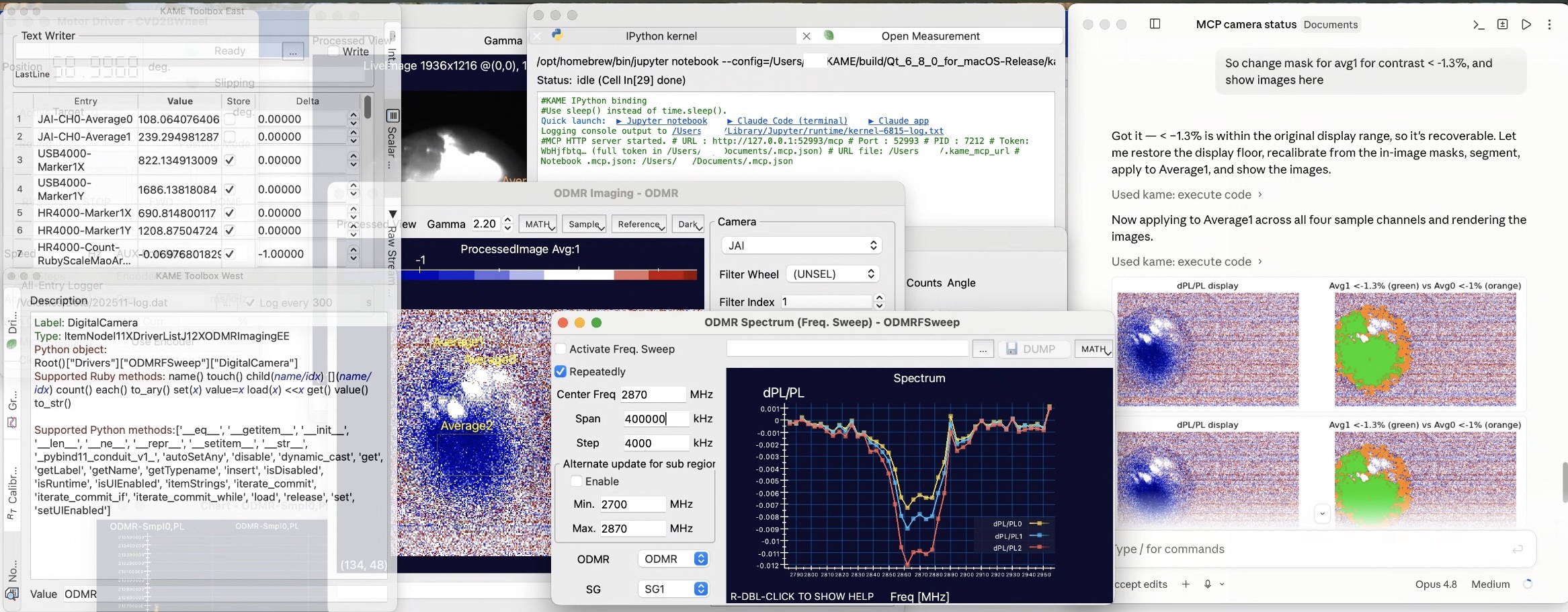}
\caption{GUI screenshot of the wide-field ODMR measurement modeled in Fig.~\ref{fig:nodetree}:
the camera image, the processed ODMR contrast map of the diamond NV centers, and
the frequency-swept spectrum. In the right panel, an AI agent (a large language
model connected through the Model Context Protocol, MCP) identifies high-contrast
fluorescence regions in natural-language dialogue with the operator,
reading the measurement state through the Snapshot API and publishing
analysis-mask updates through the Transaction API.
This workflow replaces manual region selection, which does not scale to
high-throughput wide-field ODMR.}
\label{fig:screenshot}
\end{figure*}

 The framework has been deployed mainly in two settings:
NMR (Nuclear Magnetic Resonance) experiments, where numerous threads drive asynchronous measurement---including real-time AD/DA at up to 10~MSPS---while
controlling many instruments (pulse generators, oscilloscopes, signal generators, temperature controllers, magnet power supplies), in use by multiple research groups;
and ODMR experiments for solid-state quantum sensing, where high-speed camera acquisition, image analysis, and motorized focus/optical-axis adjustment run concurrently (Fig.~\ref{fig:screenshot}).

\begin{figure}
\includegraphics[width=\columnwidth]{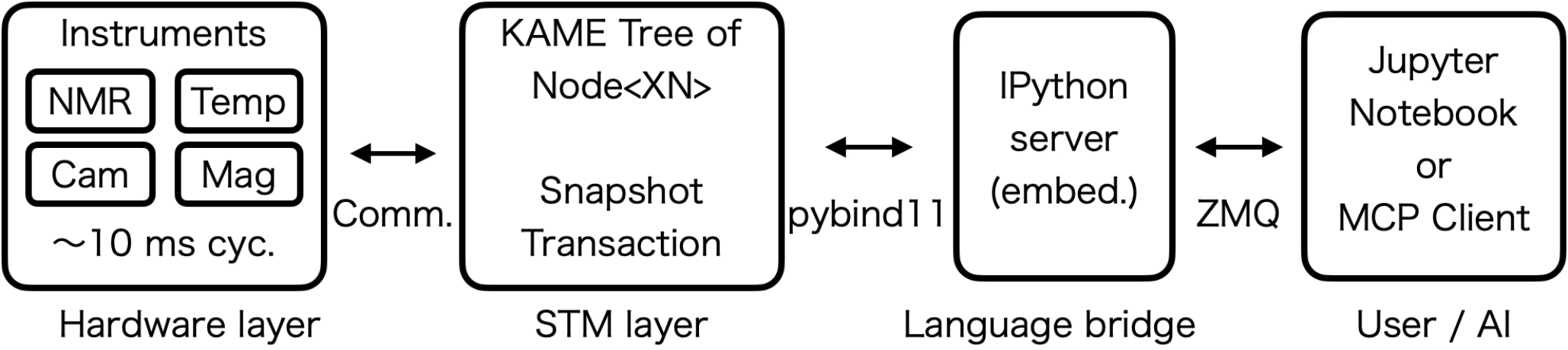}
\caption{System architecture of KAME. The central STM layer shares the tree-structured state
 among instrument drivers, native C++ clients, and external clients.
The same Snapshot/Transaction API is exposed locally through pybind11 and remotely over ZeroMQ (ZMQ);
Jupyter notebooks and AI agents connected through the Model Context Protocol (MCP)
 thus operate on the same transactional state.}
\label{fig:layer}
\end{figure}

In recent years, two properties of the STM have made it a natural substrate for external language bindings (Fig.~\ref{fig:layer}):
reference-counted \emph{lifetime management} (C++ objects survive as long as Python holds a reference)
and consistency through the same Snapshot/Transaction API.
Via pybind11,\cite{pybind11} an embedded IPython kernel\cite{perez2007ipython} with Jupyter support,
and an MCP (Model Context Protocol) server over ZMQ,
Python scripts and AI agents (large language models) read and write the node tree, 
controlling instruments and automating acquisition and analysis with the same integrity guarantees.

\section{System Design}
\label{sec:design}
The concurrency mechanism of KAME\cite{kame} is described in three stages.
Section~\ref{sec:atomic_shared_ptr} presents the lock-free \texttt{atomic\_shared\_ptr} primitive,
 introduced in 2006 and still ahead of the widely deployed standard-library implementations in this respect.
Sections~\ref{sec:stm_model}--\ref{sec:negotiate} describe the tree-structured STM adopted in 2010
---optimistic $O(1)$ snapshots, CAS (compare-and-swap)-based optimistic commits, the bundle protocol for subtree consistency, and contention management.
Section~\ref{sec:verification} reports the formal verification and the stress measurements.

\subsection{Lock-Free atomic\_shared\_ptr}
\label{sec:atomic_shared_ptr}

Realizing optimistic $O(1)$ snapshots and CAS-based commits requires shared ownership
(so that a \texttt{Snapshot} can safely retain the entire object state of a previous version)
together with lock-free operations on the shared pointer itself.
The versioning is immune to newer state as long as copy-on-write (COW) is properly conducted.
\texttt{std::atomic<shared\_ptr>}, standardized in C++20, still lacks a lock-free implementation in the major standard libraries.\cite{schafer2024lockfree}
Lock-free implementation techniques based on split reference counting were discussed in Refs.~\onlinecite{williams2012cpp,folly};
an alternative construction instead derives lock-free atomic reference-counted pointers
 from deferred-reclamation schemes, dispensing with the split count altogether.\cite{anderson2022rc}

Existing lock-free implementations illustrate the trade-offs.
Folly\cite{folly} packs a local reference count into the upper 16 bits of the pointer for a single 64-bit CAS, assuming the upper bits are free.
This packed fast path collides with any framework that uses them---wider virtual addresses, or nonzero pointer tags (e.g., Arm MTE, Android tagged heap pointers)---unless the tags are preserved.
Williams\cite{williams2012cpp} (Just::thread) applies \texttt{std::atomic} to a \texttt{counted\_ptr} structure (access\_count 32 bit + index 32 bit + pointer 64 bit = 128 bit),
 typically causing the compiler to emit a 128-bit DCAS (Double Compare-and-Swap) instruction (\texttt{cmpxchg16b} on x86\_64). 
 DCAS is substantially slower than a regular single-word CAS and lacks portability to RISC-V and embedded ARM systems.

KAME's implementation embeds a local reference counter into a Tagged Pointer
(\texttt{m\_ref}), which holds a pointer to the Ref structure in its upper
bits and the counter (0--7) in its lower 3 bits 
  (guaranteed to be zero by alignment).
As a result, using only alignment-guaranteed low bits, \texttt{m\_ref} operates with a single-word CAS that neither assumes a particular virtual-address width nor requires the upper pointer bits to be free, and needs no DCAS.
The implementation also provides \emph{intrusive counting} free from an additional heap allocation for a control block, and with better cache locality.
The Ref structure holds a global reference counter (atomic) and a pointer to the managed object. 
Under intrusive counting, the object itself serves as the Ref structure.
In the following, lock-free (in a conventional sense) protocols using \texttt{m\_ref} are described.

{acquire\_tag\_ref / release\_tag\_ref protocol}
 for an atomic read operation (\texttt{load\_shared\_}):

\begin{enumerate}
\item {acquire\_tag\_ref}: Increment the local reference counter of \texttt{m\_ref} via CAS. 
  This ``pins'' the pointer. Even if another thread replaces the pointer, the object is not deallocated as long as the local reference counter remains nonzero.
  This guarantee holds because the global counter is decremented and tested only after a successful CAS on the Tagged Pointer (i.e., only after observing the tag).
  If the local reference counter reaches its maximum (7), the thread spins and waits, although exceeding 1 is extremely rare even under stress testing.
\item {Increment the global counter by +1} (\texttt{fetch\_add}, relaxed): Acquire a reference for long-term retention.
\item {release\_tag\_ref}: Decrement the local counter by CAS. However, if another thread has replaced the pointer between steps 1 and 3, the CAS fails.
  In this case, the replacing thread has already transferred the local counter's value to the global counter, so the operation falls back to decrementing the global counter by 1.
\end{enumerate}

Procedure for a write operation (\texttt{compareAndSet} strong):
\begin{enumerate}
\item Pre-increment the global counter of the new pointer by +1
\item Verify that the old pointer matches (return failure on mismatch)
\item Transfer the old pointer's local reference counter value to its global reference counter (\texttt{fetch\_add}, relaxed)
\item Replace the pointer with the new one via CAS
\item 
  On CAS failure, subtract the transferred amount from the old pointer's global reference count and return to step 2.
\end{enumerate}
Because the caller already owns the old pointer, no pin (\texttt{acquire\_tag\_ref}) is needed inside this procedure, and
the success-path bookkeeping is folded into the transfer of step 3.

After a successful CAS, the global counter decrement (\texttt{fetch\_add}, \texttt{acq\_rel}) performs a zero-test and deallocates the object if appropriate.

A failed CAS on \texttt{m\_ref} implies, apart from transient spurious failures of the underlying primitive,
that another thread has completed a modification of the same word (a pointer replacement or a tag-count change),
so these retry loops are lock-free in the conventional sense.
However, two theoretical, perhaps not practical, liveness issues remain in this atomic\_shared\_ptr layer.
Formal liveness verification of this layer under an adversarial scheduler therefore remains open.
First, lock-freedom does not bound the retries of an \emph{individual} thread, whose CAS loop an adversarial
scheduler can starve indefinitely. In practice, under stochastic scheduling representative of practical machines,
such single-word CAS loops are known to complete in a bounded expected number of steps,\cite{alistarh2016lockfree}
and the contention window here is narrow---each attempt performs a few instructions on one shared word---so
persistent phase-aligned interference has not been empirically observed.
Second, an evident exception to lock-freedom is the rare spin at local-counter saturation, which waits for a peer's release or drain.
The current implementation manages to reduce contention and counter saturation by batching reference-count traffic:
a successful CAS may drain the packed local-reference field to zero, transferring its accumulated value to the
heap-allocated global counter in one \texttt{fetch\_add} (so the global counter may be adjusted by more than
$\pm1$), and fallback releases likewise discharge several transferred references together, reducing the
frequency of local-counter saturation.
To address these liveness concerns, arbitration at the higher layer is introduced as described below.
The weak forms of \texttt{acquire\_tag\_ref} and \texttt{compareAndSet} perform only a single attempt and report
failure to the caller, which may retry or escalate.
A reference acquired by \texttt{acquire\_tag\_ref} must eventually be released, since it participates in
lifetime protection; \texttt{release\_tag\_ref} may therefore be deferred but not discarded.
If a drain has meanwhile transferred the local count (by the current or another thread), the packed-counter
decrement is replaced by a decrement of the transferred global count.
If an operation still holds an unresolved release obligation at scope exit,
arbitration by the higher layer is performed to preserve system-wide livelock-freedom.

\subsection{STM Transaction Model}
\label{sec:stm_model}

KAME's data model represents instrument parameters and acquired data as a node tree structure.
All nodes inherit from \texttt{Node<XN>}, and reads and writes are performed through snapshots and transactions
 (explicit mutex locks are used only for synchronous communication with instruments and are never held within STM transactions).
For an isolated node, snapshot acquisition and commit reduce to a single atomic load and a single CAS (Secs.~\ref{sec:snapshot} and \ref{sec:optimistic_commit}).
Consistency across a subtree is supplied by the bundle/unbundle protocol of Section~\ref{sec:bundle}.

\subsubsection{Data Structure}
\label{sec:data_structure}

\begin{itemize}
\item {Payload}: The ``memory'' held by each node.
For a camera driver, the Payload carries image frames, exposure settings, and processed images;
for a value node (child), it holds a single scalar such as a temperature reading.
Node-specific fields are defined in subclasses of Payload.
For large or variable-size data (images, waveforms), storing the member as \texttt{shared\_ptr<const T>} enforces copy-on-write discipline at compile time.
Mutation through the pointer is prevented, and so writers must create a new object, preserving snapshot isolation.
\item {PacketWrapper}: In addition to the Payload, holds a serial number (Lamport clock\cite{lamport1978clocks}) and metadata for the bundle protocol.
\item {PacketList}: the list of child sub-packets held inside a parent's packet.
\item {Linkage}: Holds an \texttt{atomic\_shared\_ptr<PacketWrapper>}. All reads and writes are performed through this pointer. 
Also holds variables for the contention management.
\end{itemize}

\subsubsection{Optimistic $O(1)$ Snapshot}
\label{sec:snapshot}

\begin{lstlisting}[language=C++]
Snapshot<NodeA> shot(node); // one atomic load
double x = shot[node].m_x; // immutable read
\end{lstlisting}

Acquiring a \texttt{Snapshot} basically requires only a single call to \texttt{load\_shared\_} (Section~\ref{sec:atomic_shared_ptr}) of the \texttt{atomic\_shared\_ptr}.
However, the cost is not $O(1)$ when the subtree structure is in an inconsistent state.
The acquired snapshot is immutable and is unaffected by writes from other threads.
During the lifetime of the snapshot, the reference counter guarantees the survival of the object.

\subsubsection{Optimistic Transaction and CAS Commit}
\label{sec:optimistic_commit}

\texttt{Transaction} inherits from \texttt{Snapshot}; \texttt{operator[]} is a unified read-write accessor that triggers Copy-on-Write on first write.

\begin{lstlisting}[language=C++]
node.iterate_commit(
  [](Transaction<NodeA> &tr) {
    tr[node].m_x += 1; // CoW on first write
});                     // auto-retry on conflict
\end{lstlisting}

\texttt{iterate\_commit} has literal semantics; details of the family API (\texttt{iterate\_commit} / \texttt{\_if} / \texttt{\_while}) are provided in the supplementary material.

Transaction procedure:

\begin{enumerate}
\item {At construction}, take a snapshot (save the current PacketWrapper as \texttt{m\_oldpacket})
\item \textbf{\texttt{operator[]}}: Copy-on-Write one Payload inside the tree on the first write, and stamp a new serial number for the subtree
\item \textbf{\texttt{commit()}}: Single CAS on the Linkage. The CAS will fail if another thread has committed since the snapshot
\item \textbf{Retry}: \texttt{iterate\_commit} re-takes the snapshot and re-executes the lambda function
\end{enumerate}

In conventional word-based STMs such as TL2 and common TinySTM configurations, transactional reads are tracked
in a read set, and the associated version metadata is validated during execution and again at commit.
KAME instead uses the pointer identity of the immutable \texttt{PacketWrapper} as its version token.
The validation is the commit CAS itself, which compares the recorded wrapper against the currently
published one, with pointer reuse (ABA) excluded by the reference-counted lifetime.
Reads therefore require neither per-field logging, a separate version field, nor a transaction-wide validation pass.

\subsubsection{Signal Mechanism: Bridging STM and Downstream Processing}
\label{sec:signal}

The bridge between the STM and downstream processing is a necessary part of the real C++ program.
GUI rendering, signal processing, and instrument control are provided through the \emph{Talker/Listener} signal mechanism.

\textbf{Post-commit deferred notification:}
When a transaction successfully commits, a notification is issued to listeners, and the listeners receive consistent data through Snapshots.
No notification is issued for aborted attempts, so listeners never observe intermediate states of retries.

\textbf{Skipping stale events:}
In high-speed data acquisition of measurement software, the listener-side processing may not keep pace.
When the \texttt{FLAG\_AVOID\_DUP} flag is set on a listener and an identical signal already exists in the queue, it is overwritten by the new event.
This ensures that the listener always processes the most recent data, skipping stale events.
The data acquisition thread is never affected by listener delays, because it never waits for a lock.

\textbf{Affinity with STM:}
This design works well precisely because STM Snapshots are immutable.
A listener can retain a received Snapshot for as long as needed, unaffected by writes from other threads.
Since no lock on the data is ever acquired, downstream processing for GUI/script never blocks data acquisition.

\subsection{Bundle/Unbundle Protocol}
\label{sec:bundle}

While snapshots and transactions on an isolated single node are $O(1)$, 
an additional protocol is required for consistent snapshots of an entire subtree. 
KAME achieves this composability of the transactional ``memory'' through the \emph{bundle/unbundle} protocol.
When taking a snapshot of a parent node, the following protocol provides a consistent view that also includes the Payloads of child nodes.
 This is essential when the GUI simultaneously displays multiple measurement values or when an
 analysis routine or a script reads multiple 
 parameters at once.

\subsubsection{Four-Phase CAS Protocol}
\label{sec:four_phase}

\begin{figure*}
\includegraphics[width=0.7\textwidth]{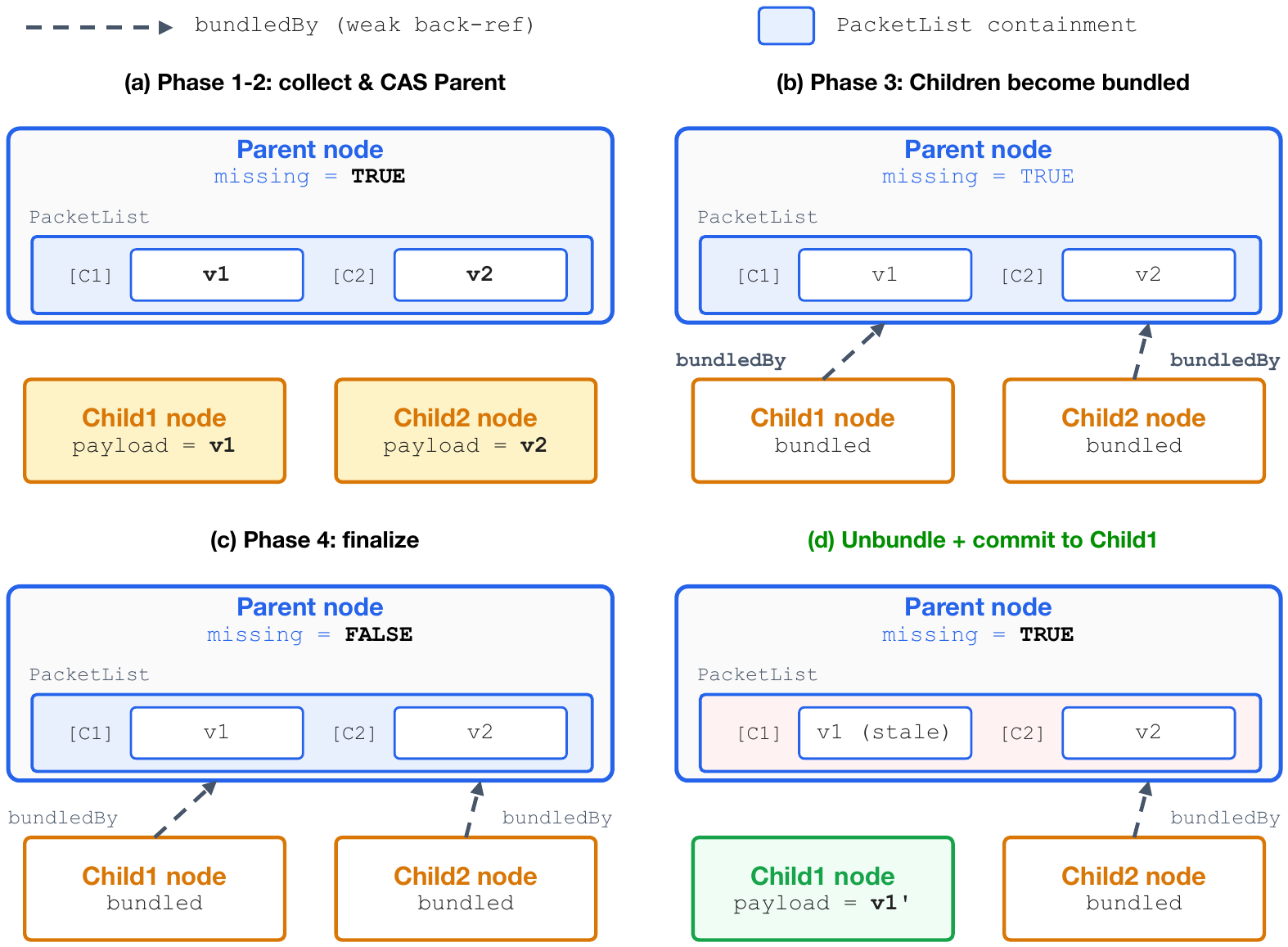}
\caption{State transitions of the four-phase bundle/unbundle protocol, drawn as
the actual pointer graph. The Parent's \texttt{PacketList} {contains} a snapshot
of each child sub-packet (nested boxes). A bundled
child's Linkage holds a \texttt{BundledRefWrapper} whose \texttt{packet} is
\texttt{Null} and whose \texttt{bundledBy} is a weak back-reference up to the
Parent (dashed arrow); a snapshot of that child therefore follows
\texttt{bundledBy} to the Parent and reads \texttt{Parent.PacketList[child]}.
On commit (unbundle, panel d), the committing child is restored to a regular packet-holding \texttt{PacketWrapper}
while the Parent's sub is kept for recovery.}
\label{fig:bundle}
\end{figure*}

The protocol consists of four phases (Fig.~\ref{fig:bundle}):
Phase~1 (Collect) reads the PacketWrapper of each child node.
Phase~2 (CAS Parent) installs a new PacketWrapper (missing=TRUE) on the parent's Linkage.
Phase~3 (CAS Children) replaces the Linkage of each child with a BundledRefWrapper (a back-reference to the parent).
Phase~4 (Finalize) updates the parent's Linkage to missing=FALSE, completing the process.

\begin{itemize}
\item \textbf{missing flag}: Indicates that the child slot in the parent packet is stale (not up to date).
  This flag is set to TRUE during Phases 2--4.
\item \textbf{BundledRefWrapper}: Indicates that a child node is bundled.
  The child's Linkage holds a back-reference to the parent, forwarding snapshot requests to the parent.
\end{itemize}

\subsubsection{Unbundle (at Commit Time)}
\label{sec:unbundle}

When committing to a bundled child node, one possible strategy is to always copy-on-write the entire data from the bundle root and CAS it.
However, in that case the commit would conflict with concurrent commits to other, even unrelated, parts of the subtree.
It is undesirable for data acquisition commits by one instrument driver to conflict with commits by another instrument driver every time. 
Therefore, in this STM, the child is unbundled at commit time.

\begin{enumerate}
\item Update the parent's Linkage via CAS (setting missing=TRUE). This marks the slot as stale.
\item Restore the child's Linkage from a BundledRefWrapper to a regular PacketWrapper containing the new Payload via CAS
\end{enumerate}

In a three-level tree (Grand $\rightarrow$ Parent $\rightarrow$ Child), a two-stage unbundle is required.
Subsequent transactions on the child node are $O(1)$ until the next bundle operation.

\textbf{Safety of unbundle.}
Unbundle temporarily breaks the ancestors' subtree consistency:
 the child receives a new independent packet (with the committed Payload), while the ancestors' nested sub-slots still hold the old copy.
However, the ancestors' missing flags are set \emph{before} the child's Linkage is modified (step~1 above), 
so any concurrent snapshot of an ancestor observes the stale slot and triggers a re-bundle, which recovers consistency on demand.
the invariant that every completed snapshot reflects a single committed
version (SnapshotConsistency; supplementary material) holds across all unbundle interleavings in the TLA+ model (Section~\ref{sec:verification}).

\subsubsection{Extension: Multi-Parent Nodes}
\label{sec:multiparent}

In KAME's tree structure, a child node normally has a single parent, but \emph{multi-parent nodes} 
(a configuration in which a single child node has multiple parents, forming a DAG) are also supported.
In KAME, this is used for nodes representing scalar quantities output by instrument drivers: 
by having both the instrument driver itself and a list node for the scalar quantity overview as parents, two modes of management are provided.

When a Snapshot/Transaction is executed against one parent A, 
the tree structure from the other parent B is unbundled, temporarily completing parent A's tree.

Even without multi-parent nodes, taking a Snapshot of a level above the target node within a transaction scope triggers a bundle operation and causes the transaction to inevitably fail. 
With multi-parent nodes, a Snapshot involving the seemingly unrelated parent B can also cause failure, which requires caution.
In all cases, the result is a CAS failure (a liveness issue), not data corruption.
Snapshot/Transaction provides contextually intuitive coding, but these behaviors are non-intuitive and error-prone in practice.
It is necessary to take Snapshots/Transactions at the appropriate tree level.

\subsubsection{Extension: Dynamic Tree Structure Changes}
\label{sec:dynamic_tree}

In measurement software, it is necessary to dynamically add, remove, and swap the order of child nodes at runtime
 (e.g., connecting a new instrument, generating analysis tools from a script).
KAME's STM supports transactional node operations.

\begin{lstlisting}[language=C++]
parent.iterate_commit_if(
  [&](Transaction<NodeA> &tr) -> bool {
    if(!parent.insert(tr, child, true))
      return false; // online insertion
    tr[*child].m_x = 42; // immediate access
    return true;
});
\end{lstlisting}

\begin{itemize}
\item \textbf{\texttt{insert(tr, child, true)}}: The third argument \texttt{true} (online insertion) reflects the child node in the live tree immediately, without waiting for the transaction to commit.
  This allows read/write access via \texttt{tr[*child]} within the same transaction.
  If \texttt{true} is omitted, the child node becomes visible only after the commit.
\item \textbf{\texttt{release(tr, child)}}: Removes a node. Must be called after confirming existence via \texttt{Snapshot::isUpperOf(*child)} (to prevent double release during shutdown).
\item \textbf{\texttt{swap(tr, child1, child2)}}: Swaps nodes.
\end{itemize}

These operations internally involve modifications to the PacketList, and conflicts with other threads 
are detected by CAS at commit time.
In the case of online insertion/release, a multi-phase CAS is used. 
Correctness of concurrent node additions and deletions is verified in
Sections~\ref{sec:layer2} and by the stress tests listed in the
supplementary material.

\subsection{Contention Management: Livelock-Free Negotiation}
\label{sec:negotiate}

In the STM literature, the module responsible for resolving conflicts between concurrent transactions is 
called a contention manager.\cite{scherer2005contention,guerraoui2005greedy}
In KAME STM, the management is essential not only for transactions but for bundle operations and snapshots. 
Negotiation is activated when the CAS of commit (Section~\ref{sec:optimistic_commit}), 
bundle/unbundle (Sections~\ref{sec:four_phase}--\ref{sec:unbundle}), or online insertion/release (Section~\ref{sec:dynamic_tree}) fails.
In a relatively simple implementation,
proportional backoff is performed based on the timestamp at the start of the initial transaction or snapshot,
giving priority to long-running transactions
 (in the current implementation the wait is realized as brief timed waits, with randomized jitter and notify-based early wake-ups, rather than as a pure timed sleep).
This suppresses livelock under high write contention, and
the effectiveness and necessity of the backoff are demonstrated in the stress measurements presented below.

The negotiation protocol is not wait-free but lock-free (guaranteeing system-wide progress).
To guarantee progress of individual transactions, each Linkage carries a start-time and thread-ID stamp that identifies the oldest active transaction or snapshot (oldest TX).
When the oldest TX enters definite contention, its stamp is registered in the Linkage variable and other threads yield their CAS operations on that node,
 entering a per-linkage priority mode.
If a linkage is already stamped, the oldest timestamp takes precedence.
When the stamping TX finishes, it clears its own stamp with a single CAS (atomically checking that the stamp is still its own);
a stamp that an even-older TX has meanwhile overwritten is left for that TX to clear.
By this mechanism the oldest TX is guaranteed to complete within a finite number of CAS operations, and TXs on independent subtrees remain unaffected.
Because each thread runs at most one active transaction, at most $T{-}1$ older transactions precede any given one under this arbitration.

\section{Formal Verification and Testing}
\label{sec:verification}

\subsection{Verification Strategy: A Two-Layer Approach}
\label{sec:two_layer}

Prior work on the formal verification of concurrent programs includes CDSChecker by Norris \& Demsky (2013), 
an earlier model checker that
exhaustively explores executions of programs using C11/C++11 atomics
under the C/C++ memory model.\cite{norris2013cdschecker}
 GenMC by Kokologiannakis et al.\ (2019)\cite{kokologiannakis2019genmc} is a stateless
model-checking framework that is parametric in the choice of weak memory
model. Under its assumptions, it explores each consistent
execution exactly once while avoiding inconsistent executions and
redundant exploration paths.
 TLA+ by Lamport (2002)\cite{lamport2002tla} is a formal specification language for modeling concurrent and distributed systems, 
 enabling automated verification via the TLC model checker.\cite{yu1999tlcmc} 
 By combining these tools, we verify lock-free data structures from both the protocol-logic and memory-model perspectives.

Verifying the entire KAME STM stack with a single model is impractical due to state-space explosion.
We therefore adopted a strategy of partitioning the system into two layers at different levels of abstraction,
 verifying each layer with the appropriate tool: Layer~1 (atomic\_shared\_ptr) is checked with TLA+ 
 and GenMC (RC11), and Layer~2 (bundle/unbundle + commit) with TLA+ (TLC).
The Layer~2 abstracts the \texttt{atomic\_shared\_ptr} verified in Layer~1 as a correct atomic register. 
This two-layer verification architecture is summarized in the supplementary material.
As a key difference from the C++ implementation, 
the TLA+ models are conservative without the fast-path optimizations active only for short time periods.

The TLA+ models, the GenMC (C11) test programs, the 2/3-level static tree C++ stress tests, and the related scripts were developed with the assistance of a large language model
 (Claude Opus 4.6--4.8, Anthropic), 
 used as a coding assistant under the author's direction and chosen for its ability to draft TLA+ specifications and test code from the C++ source.
 All models, invariants, and scripts were reviewed by the author, and every verification result reported here is the direct output of TLC/GenMC runs.

\subsection{Layer 1: atomic\_shared\_ptr (TLA+ + GenMC)}
\label{sec:layer1}

Layer~1 (\texttt{atomic\_shared\_ptr}) was verified with both TLA+ for the reference-counting protocol logic,
 and GenMC for the C11/RC11 memory ordering (\texttt{tests/cds\_atomic\_shared\_ptr/}).
For full details of this section (complete model, invariants, verification configuration, etc.),
see the supplementary material.

\subsection{Layer 2: Bundle/Unbundle Protocol (TLA+)}
\label{sec:layer2}

\subsubsection{Model and Verified Configurations}
\label{sec:layer2_model}

\begin{figure}[htbp]
\centering
\begin{tabular}{c@{\hspace{3mm}}c@{\hspace{3mm}}c@{\hspace{3mm}}c}
\begin{tikzpicture}[node distance=7mm]
  \node[tnode] (p) {P};
  \node[tnode] (c1) [below left=6mm and 3mm of p] {C1};
  \node[tnode] (c2) [below right=6mm and 3mm of p] {C2};
  \draw[tedge] (p) -- (c1); \draw[tedge] (p) -- (c2);
\end{tikzpicture}
&
\begin{tikzpicture}[node distance=6mm]
  \node[tnode] (g) {G};
  \node[tnode] (p) [below=of g] {P};
  \node[tnode] (c1) [below left=6mm and 3mm of p] {C1};
  \node[tnode] (c2) [below right=6mm and 3mm of p] {C2};
  \draw[tedge] (g) -- (p); \draw[tedge] (p) -- (c1); \draw[tedge] (p) -- (c2);
\end{tikzpicture}
&
\begin{tikzpicture}[node distance=7mm]
  \node[tnode] (p) {P};
  \node[tnode] (c1) [below left=6mm and 3mm of p] {C1};
  \node[tnode] (c2) [below right=6mm and 3mm of p] {C2};
  \draw[tedge] (p) -- (c1);
  \draw[dedge] (p) -- node[right]{ins/rel} (c2);
\end{tikzpicture}
&
\begin{tikzpicture}[node distance=6mm]
  \node[tnode] (g) {G};
  \node[tnode] (p) [below=of g] {P};
  \node[tnode] (c1) [below left=6mm and 3mm of p] {C1};
  \node[tnode] (c2) [below right=6mm and 3mm of p] {C2};
  \draw[tedge] (g) -- (p); \draw[tedge] (p) -- (c1);
  \draw[dedge] (p) -- node[right]{ins/rel} (c2);
\end{tikzpicture}
\\[1mm]
{\scriptsize (a) 2-level} & {\scriptsize (b) 3-level} & {\scriptsize (c) 2-level dyn.} & {\scriptsize (d) 3-level dyn.} \\[3mm]
\begin{tikzpicture}[node distance=6mm]
  \node[tnode] (r) {R};
  \node[tnode] (a) [below left=6mm and 4mm of r] {A};
  \node[tnode] (b) [below right=6mm and 4mm of r] {B};
  \node[tnode] (c) [below=11mm of r] {C};
  \draw[tedge] (r) -- (a); \draw[tedge] (r) -- (b);
  \draw[tedge] (a) -- (c); \draw[tedge] (b) -- (c);
\end{tikzpicture}
&
\begin{tikzpicture}[node distance=6mm]
  \node[tnode] (r) {R};
  \node[tnode] (a) [below=of r] {A};
  \node[tnode] (c) [below=of a] {C};
  \draw[tedge] (r) -- (a); \draw[tedge] (a) -- (c);
  \draw[tedge] (r) to[bend left=55] (c);
\end{tikzpicture}
&
\begin{tikzpicture}[node distance=6mm]
  \node[tnode] (g1) {GN1};
  \node[tnode] (p1) [right=9mm of g1] {P1};
  \node[tnode] (p2) [below=8mm of $(g1)!0.5!(p1)$] {P2};
  \draw[tedge] (g1) -- (p2); \draw[tedge] (p1) -- (p2);
\end{tikzpicture}
&
\begin{tikzpicture}[node distance=6mm]
  \node[tnode] (g1) {GN2};
  \node[tnode] (p1) [right=9mm of g1] {P1};
  \node[tnode] (p2) [below=8mm of $(g1)!0.5!(p1)$] {P2};
  \draw[tedge] (p1) -- (p2);
  \draw[dedge] (g1) -- node[left]{migr.} (p2);
\end{tikzpicture}
\\[1mm]
{\scriptsize (e) 4-node DAG} & {\scriptsize (f) self-collision} & {\scriptsize (g) external} & {\scriptsize (h) ext.\ migration} \\
\end{tabular}
\caption{Parent--child topologies verified by the Layer~2 TLA+ models. 
(a)--(d) tree structures; (e)--(h) multi-parent (DAG) structures where a child is shared by two parents.
 Solid = static edge; dashed (ins/rel, migr.) = runtime insert/release/migration.}
\label{fig:topologies}
\end{figure}
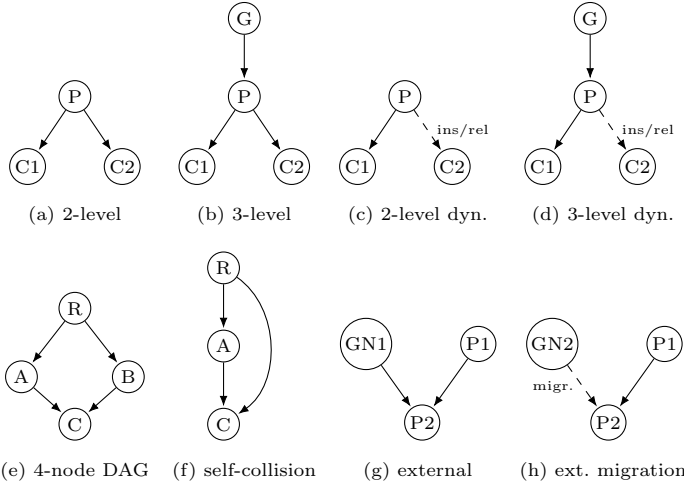

The bundle/unbundle protocol was modeled in configurations for
a \emph{2-level tree} (Parent $\rightarrow$ \{Child1, Child2, \ldots\}) and a \emph{3-level tree} 
(Grand $\rightarrow$ Parent $\rightarrow$ \{Child1, Child2, \ldots\}). Beyond these tree structures, 
multi-parent (DAG) topologies in which a child is shared by two parents were also verified. 
Figure~\ref{fig:topologies} shows all verified topologies; 
solid arrows are static structural edges and dashed arrows (labelled) are runtime \texttt{insert}/\texttt{release}/migration operations in the dynamic models.

The verification covers (i)~the 4-phase bundle protocol (collect $\rightarrow$ CAS parent $\rightarrow$ CAS children $\rightarrow$ finalize),
 (ii)~unbundle at commit time (CAS parent sets missing=TRUE $\rightarrow$ CAS child restores its regular packet),
(iii)~interference from concurrent snapshot/commit operations, and (iv)~the single-node commit optimization.
For the dynamic models, it also covers (v)~concurrent insertion and deletion of a child.

\textbf{Verification patterns (corresponding C++ code):}
Threads are divided into two roles of \emph{parent-scope commit} (targeting the parent in 2-level, or the grandparent in 3-level),
 and \emph{leaf-direct commit} (committing to each child node individually).
The following is an example of a 2-thread verification.

\begin{lstlisting}[language=C++]
// 3-level: one iteration of a thread
if (thread_performs_grandparent_scope_commit)
  grand.iterate_commit([&](Transaction<NodeA> &tr) { // snapshot(grand) -> recursive bundle
        tr[child1].m_x += 1;
        tr[child2].m_x += 1;
  });
if (thread_performs_leaf_direct_commit)
  for (auto& child : {child1, child2})
      child.iterate_commit( // leaf commit
        [&](Transaction<NodeA> &tr) {
          tr[child].m_x += 1;
      });
\end{lstlisting}

Upon iteration completion, each child payload's \texttt{m\_x} can be compared with each thread's iteration count.
 This serves as a terminal invariant for verification. 
Each verification configuration partitions the state space by contention path 
by constraining commit targets. Representative configurations for 3 threads are shown in Table~\ref{tab:configs}.
\begin{table*}
\caption{Representative 3-thread verification configurations for Layer~2.}
\label{tab:configs}
\begin{tabular}{llp{1.8cm}p{6cm}}
\toprule
Config. & Parent-scope commit & Leaf-direct commit & Primary contention path verified \\
\midrule
A & Thread 2 only & Threads 1, 3 & Bundle/unbundle concurrency (1 thread bundling while 2 threads unbundle) \\
B & Threads 2, 3 & Thread 1 & CAS tag contention (2 threads competing for parent CAS) \\
C & All threads & None & Bundle collect 3-way contention (leaf-direct commit path not executed) \\
\bottomrule
\end{tabular}
\end{table*}

\subsubsection{Liveness Checking}
\label{sec:layer2_liveness}

In the STM layer, artificial state-space constraints are not used because the state space is structurally finite owing to the negotiation protocol.
Whereas the C++ implementation delays the transition to the 
per-linkage priority slot mode in consideration of priorities and time slices,
 the TLA+ model performs this transition eagerly to capture the essence of the protocol.
Considering that the Lamport serial increases monotonically, if verification terminates within a finite number of states,
 livelock freedom along paths with serial increments is proven.
In 2- and 3-thread configurations, \texttt{EventuallyAllDone} was exhaustively explored as a temporal property,
 providing machine-verified liveness.
For 2-thread configurations, the exploration completed for all static and dynamic models---both safety and
 liveness, except that the 3-level dynamic model with release completed safety only.
For 3-thread configurations, verification was limited due to computational cost constraints:
 in the static models, the all-root contention pattern (Config~C in Table~\ref{tab:configs}), Bundle collect 3-way contention,
  passed exhaustive exploration of both safety and liveness (137M states for 2-level; 541M states for
 3-level), and dynamic configurations that permit node insertion and prohibit release
 also passed both (two configurations splitting insert + parent + leaf roles over three threads, each with 100K--150K states).
  General static 3-thread configurations including leaf-only threads do not complete within practical time limits
   due to the nonlinear cost scaling of TLC with state count. We therefore combine these limited 3-thread configurations with 
   the complete 2-thread exploration and empirical evidence from C++ stress tests (Section~\ref{sec:benchmarks}, up to 128 threads). 
   For exact definitions of each configuration, state counts, execution times, and details of
    the 3-level tree and dynamic insert/release models, see the supplementary material.

\subsubsection{Generalization and Verified Scope}
\label{sec:layer2_generalization}

The exhaustively verified cases (2/3-level $\times$ static/dynamic
$\times$ 2--3 threads $\times$ 2 children) cover the structurally distinct interference patterns of the {tree} protocol---by argument for depth and width, and by projection evidence for the
  thread count, as follows.
We establish a {structural reduction} in three parts: an inductive argument over depth, a permutation-symmetry argument over width,
 and, for the thread count, projection evidence of saturation that we state as a conjecture rather than a proved cutoff.
The first two reduce correctness on any tree instance to the small fixed instance (depth 3, width 2) that the TLA+ model covers.
The reduction rests on one structural property: every bundle and unbundle operation is a {recursive traversal that applies the same local operation at each node},
 and each per-node step is uniform in both depth and width.
 (a)~\textit{Depth}: unbundle reads the whole root-to-target chain (a recursive fold that mutates nothing. Stale reads are rejected by the commit's pointer-identity check), 
 then writes one \texttt{compareAndSet} per ancestor. Bundle descends symmetrically. 
 Each per-level step is identical and has the same footprint (the node, its immediate fold-neighbour, and the root anchor),
  independent of depth. The 3-level instance instantiates all three node roles (root, internal, leaf);
   a height-$(H{+}1)$ tree adds one interior node with the identical step.
   Taking the exhaustively checked $H{=}3$ instance as the base case, safety carries over to any deeper $H$ by induction.
     (b)~\textit{Width}: additional children are processed sequentially in Phase~3,
     each per-child CAS being symmetric to the verified two-child case,
     while Phase~2 swaps the whole child \texttt{PacketList} in a single CAS, independent of the child count.
     This width argument relies on permutation symmetry among children. 
     (c)~\textit{Threads}: 
     To test whether additional threads ($T > 2$, $T$: thread count) create new behaviour, 
     we project each reachable state onto the tree's structural shape (per node: whether it holds its own packet or is bundled into a parent,
      and whether its slots are up to date), discarding thread identities and data values. 
     None of the safety invariants reads a thread identifier or serial number.
      For the 2-level static model, 
      the tree can take only six distinct configurations (the stages of the bundle/unbundle protocol of Fig.~\ref{fig:bundle},
       including the two mid-Phase-3 states in which one child is bundled before the other), 
       and this set is identical at $T{=}2$ and $T{=}3$ over the complete $T{=}3$ exhaustion,
        while the raw state count grows by three orders of magnitude.
         (The 3-level $T{=}3$ exhaustion was 541M states. Dumping and projecting this was beyond current HPC capacity.)
          We therefore claim safety for arbitrary finite depths and widths under the verified commit roles (root- and leaf-targeted transactions),
          and present $\forall T$ as a conjecture with strong supporting evidence---the projection discards thread-local control state,
          so saturation is evidence, not a cutoff proof. The details and the saturation data are in the supplementary material.

\textbf{Liveness} (livelock-freedom) can be drawn without a depth/thread cutoff, but with the assumptions the TLA+ models encode: the transaction bodies are finite, the traversed structure is finite and acyclic, and scheduling is weakly fair.
Under the oldest-first priority, the oldest active transaction is never permanently blocked by a younger one,
 and any disturbance is itself a peer's successful commit,
 that is, genuine system-wide progress.
 The oldest eventually runs undisturbed and completes, and the next-oldest continues.
  This is the same oldest-wins argument that establishes starvation-freedom for priority contention managers.\cite{guerraoui2005greedy}

The multi-parent (DAG) topologies
 (Fig.~\ref{fig:topologies}e--h) do not generalize by the tree induction alone. 
 When a child is shared by two parents, the \texttt{bundledBy} chain is no longer a unique path, and a bundle on one parent can interact with 
 a bundle/unbundle on the other through the shared child.
  This interaction is localized to the shared child, so we verify the specific \emph{acyclic} multi-parent structures that the production code can actually construct.
   A verified fragment composed with trees (trees above its parents or below its shared child) would inherit safety by the same uniformity argument:
    operations on a shared child have the same $\{\text{self}, \text{parent}, \text{root anchor}\}$ footprint as tree operations, 
    the cross-parent coupling being only the root-reachability check that the Phase~4 gate verifies. 
    We do \emph{not} claim coverage of arbitrary acyclic DAGs (many shared children or parents),
   and graphs containing cycles are outside the verified scope entirely.

A summary of the verification results across all layers, as well as complete model text, invariant definitions, and
state-count tables, is provided in the supplementary material.
The 3-thread configurations for Layer~1 and for Layer~2 were executed on the ISSP supercomputer (Ohtaka).

\subsection{C++ Stress Tests}
\label{sec:stress_tests}

Formal verification guarantees correctness on abstract models, but cannot detect bugs
arising from C++ optimization fast paths, memory layout, or hardware-specific behavior.
C++ stress tests complement this gap. 
All stress tests and GenMC runs completed with no safety violation.
The complete list of C++ stress tests is provided in the supplementary material.

\subsection{Stress Measurements: Livelock Reproduction and Removal}
\label{sec:benchmarks}

Sustained throughput as a function of thread count $T$ was measured on three machines,
including a measurement-software production environment in our laboratory
 and HPC nodes extending the evaluation to higher core counts and contrasting processor architectures.
A run is classified as livelocked when no child update completes during its stress window time (3 sec.) although the threads keep issuing CAS attempts;
a configuration is marked \textit{LL} when all three stress runs are livelocked.
The purpose of these measurements is to reproducibly trigger the livelock of the legacy protocol and to confirm that, under the same conditions,
 it no longer occurs with the revised negotiation protocol. Because finite stress runs cannot establish absence under every schedule,
 this is implementation validation complementary to the model-checked liveness of Section~\ref{sec:layer2_liveness}.
Secondarily, the measurements quantify the overhead of the protocols. They are not intended as a performance benchmark against the other approaches.
The detailed specifications of the measurement environments are listed in the supplementary material.

\textbf{Three variants compared:}

\begin{itemize}
\item \textbf{v0 (legacy)}: The complete source tree at git \texttt{1d49b07c} (early 2026). 
This is the proportional-wait backoff strategy that has been in production for 16 years.
\item \textbf{no\_backoff} (built with \texttt{-DKAME\_STM\_DISABLE\_BACKOFF=1}): 
Current code with the backoff layer completely bypassed. No sleep is performed, and the variant is prone to livelock.
\item \textbf{oldest-first} (named \texttt{full} in the benchmark scripts and data files): 
current code with per-linkage priority mode with livelock-free mechanism,
and adaptive anti-contention threading enabled.
\end{itemize}

Measurements were run under two contention profiles similar to the above TLA+ models:
\texttt{transaction\_payload\_integrity\_mixed\_test} on a 2-level tree (2L; Parent $\rightarrow$ Child1,Child2,\ldots,Child$_T$)
and \texttt{transaction\_payload\_integrity\_3level\_mixed\_test} on a 3-level tree (3L; Grand $\rightarrow$ Parent $\rightarrow$ Child1,Child2,\ldots,Child$_T$).

Each of the $T$ threads transactionally increments the payload counter of its corresponding single Child node, but once every CR (cross-transaction ratio) iterations, 
it commits as a Parent or Grand transaction that increments all $T$ Child nodes.
That is, CR=$\infty$ means all operations are leaf-only and fully disjoint; 
CR=2 means 1/2 are parent/grand scope (alternating bundles); CR=10 means 1/10 are parent/grand (the remaining 9/10 are leaf-only), a regime prone to livelock due to \emph{parent-induced leaf rollbacks}.

We report \emph{child updates/s (cu/s)}: the total number of child-node payload writes, divided by the run time. 
A single subtree-wide (parent/grand) commit atomically updates all $T$ children, so for CR${<}\infty$ each such commit contributes $T$ to cu/s;
 cu/s therefore exceeds the raw commit rate, and only in the leaf-only regime (CR=$\infty$) is cu/s identical to commits/s. 
 Note that this workload is designed to {scale contention with $T$}: every subtree-wide commit conflicts with all $T$ child threads, 
 so the conflict scope grows with the thread count. Unlike fixed-size STM benchmarks (e.g.\ STAMP\cite{minh2008stamp}, Synchrobench), 
 whose shared structure is independent of $T$, 
 our workload is therefore an \emph{adversarial livelock stress test} rather than a throughput-scaling benchmark. 
 Genuine $T$-scaling is meaningful only at CR=$\infty$. The CR=2/10 sweeps measure resilience to contention that intensifies with $T$.

The following observations focus on the heavy-contention regime at CR=10 (Fig.~\ref{fig:scaling}).
Full sweeps for CR=$\infty$/CR=2 are provided in the supplementary material.

\begin{figure*}
\includegraphics[width=0.7\textwidth]{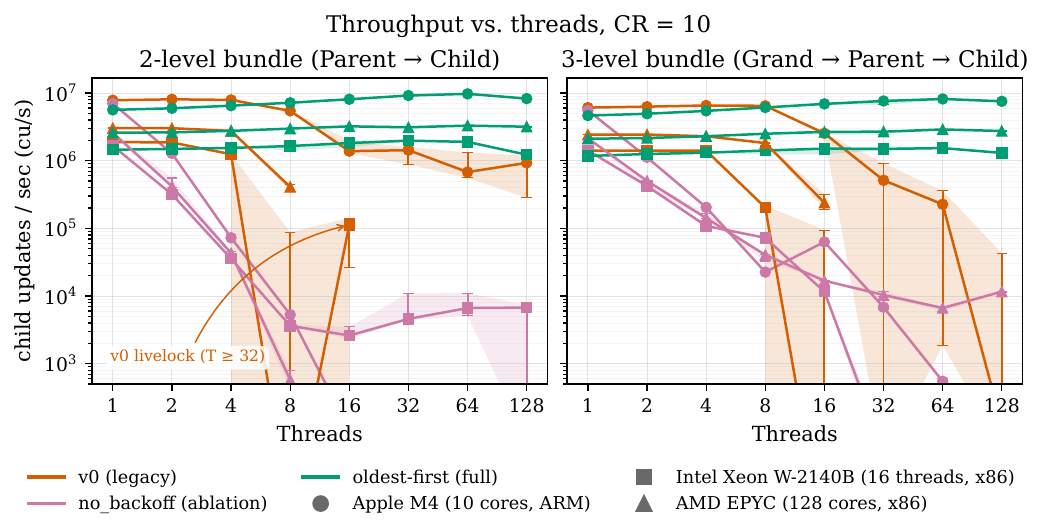}
\caption{CR=10 (heavy mixed contention) throughput vs.\ thread count on log-log axes. 
Three platforms (Apple M4 $\circ$, 10-core ARM heterogeneous; Intel Xeon W-2140B $\square$, 8C/16T x86 TSO homogeneous; AMD EPYC $\triangle$, 128-core x86 TSO homogeneous NUMA) $\times$ 2 workloads (left: 2-level, right: 3-level). Color = variant (v0 orange, no\_backoff purple, oldest-first green), line = median, fill = min--max range.
On each platform, all variants share an identical injected allocator, so the comparison isolates the STM algorithm.}
\label{fig:scaling}
\end{figure*}

\textbf{Key observations}:

\begin{enumerate}
\item \textbf{v0 showed livelock on strong-memory x86}: On iMac Pro (Xeon W-2140B), 
v0 reaches 1.25~M~cu/s at $T$=4 in CR=10 2L but collapses at $T$=8 (median 5~cu/s, range \texttt{[0, 88K]}),
 entering hard livelock with all 3 runs at 0~cu/s for $T\geq$32 in both 2L and 3L. 
 This demonstrates that the legacy proportional-wait backoff falls into a lock-step re-execution cycle under truly parallel CAS on the strong-memory x86 TSO model.
  On Apple M4 (weak-memory ARM), v0 degrades severely but does not reach complete 0: CR=10 2L $T$=128 yields 0.93~M~cu/s (down from 7.8~M at $T$=1).
   The M4's heterogeneous core configuration (P$\times$4 + E$\times$6) or scheduling may introduce natural jitter that partially breaks the lock-step cycle.

\item \textbf{Resolution through per-linkage priority mode}: The oldest-first variant sustains 1.23~M~cu/s at 2L CR=10 $T$=128 and 1.30~M~cu/s at 3L CR=10 $T$=128 on iMac Pro,
 fully eliminating livelock in all measured conditions. On M4, it sustains 8.29~M~cu/s at 2L CR=10 $T$=128 and 7.56~M at 3L. 

\item \textbf{Catastrophic collapse without backoff}: no\_backoff collapses to near-zero throughput (thousands of cu/s down to zero) from $T \geq$8 onward on both platforms
 (iMac Pro CR=10 2L $T$=128: $\sim$7~kcu/s; M4: 0~cu/s). This confirms that the backoff layer itself is essential.

\item \textbf{128-core NUMA}: The oldest-first variant maintains a median of 3.18~M~cu/s up to $T$=128 in 2L CR=10 on Ohtaka (the AMD EPYC platform) (range \texttt{[3.15M, 3.19M]}), 
while v0 and no\_backoff both livelock. For 3L CR=10, it achieves 2.88~M~cu/s at $T$=64 and a median of 2.78~M~cu/s at $T$=128 (range \texttt{[2.77M, 2.80M]}). 
Within the measurement scope of this study, no residual livelock cells exist for the oldest-first variant.
\end{enumerate}

The complete table of CR=10 median throughput across all platforms, variants, and thread counts, 
together with a comparison of the three variants at $T$=128 with CR=2, is provided in the supplementary material.

\begin{figure}
\includegraphics[width=\linewidth]{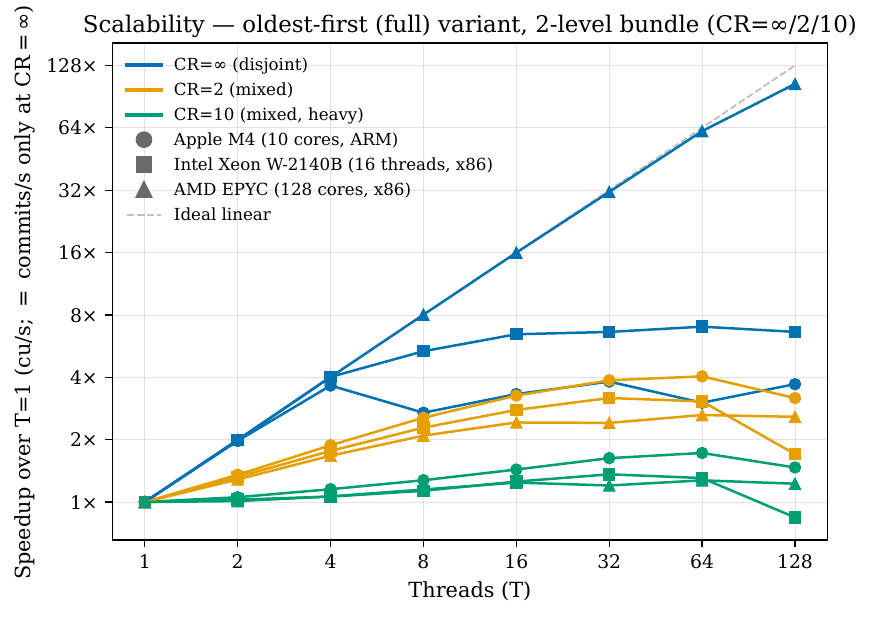}
\caption{Speedup relative to $T$=1 for the oldest-first variant, 2-level workload (log-log).
 Color = contention level (CR=$\infty$ disjoint blue, CR=2 mixed orange, CR=10 mixed green), 
 marker shape = platform (Apple M4 $\circ$, heterogeneous ARM 10c; Xeon W-2140B $\square$,
 x86 16T; AMD EPYC $\triangle$, x86 128c).
Dashed silver = ideal linear for comparison with the CR=$\infty$ case. 
}
\label{fig:speedup}
\end{figure}

Figure~\ref{fig:speedup} shows speedup relative to $T$=1 for the oldest-first variant (2L). 
Genuine throughput scaling is meaningful only at CR=$\infty$ (leaf-only).
There Ohtaka reaches 104$\times$ at 128 cores and iMac Pro 6.5$\times$ at 16 hardware threads, 
 confirming that the lock-free leaf path scales near-ideally when transactions are disjoint. 
 The CR=2 and CR=10 curves must be read with care. Because the workload scales contention with $T$
  (each subtree commit conflicts with all $T$ children) and cu/s counts $T$ child updates per such commit, 
  these curves do {not} represent commit-rate scaling. Their significance is instead \emph{stability}: 
  the oldest-first variant sustains a positive coherent-update rate at every $T$,
  whereas v0 collapses to livelock and no\_backoff to near-zero (Fig.~\ref{fig:scaling}). 
  The raw commit rate under CR=2/10 in fact declines with $T$, exactly as expected when the conflict scope grows 
  with the thread count.

Across all measured (test, CR, $T$) configurations, the oldest-first variant exhibited no livelock.

\section{Discussion}
\label{sec:discussion}

\subsection{Measurement Software Design and Operational Track Record}
\label{sec:track_record}

In measurement software, the essential requirements are that data acquisition threads are executed in accord with their priority or that fairness is guaranteed,
 and that acquired data can be readily and reliably passed to analysis routines.

For the {data acquisition threads} (running, e.g., every 10~ms), practical measurement configurations typically assign one primary writer to each node, 
so write contention is relatively rare.
On the uncontended fast path, publication completes with one successful CAS;
Retries are normally caused by a concurrent update to the same published node state or by a bundle/unbundle transition affecting it.
After a successful commit, an asynchronous signal is emitted to propagate the acquired information to analysis routines and instrument control I/O.
Snapshots carry acquisition timestamps, so subsequent processing can correlate and order observations from multiple instruments by acquisition time rather than by signal-delivery order.
The \emph{GUI threads} primarily consume the immutable Snapshots delivered after acquisition commits, so rendering never holds a lock on the live tree;
when the GUI takes new Snapshots, the event-elision (Section~\ref{sec:signal}) and negotiation (Section~\ref{sec:negotiate}) mechanisms
reduce redundant notifications and give acquisition transactions priority within the arbitration protocol.
The \emph{script threads} access C++-managed shared state through the Python Snapshot and Transaction bindings, without exposing unsynchronized mutable references.
Transactions driven over inter-process communication can be long-lived, where oldest-first arbitration prevents their indefinite starvation under the assumptions of Section~\ref{sec:layer2_generalization},
so external operations of unbounded duration should be designed with a shorter transaction if strict atomic publication over many nodes is not needed: take a Snapshot, perform the external work, and publish the result through a short validating Transaction.

KAME has been in use since its initial release in 2004 across diverse measurement environments, including electrical resistivity measurements,
NMR (10~MSPS I/O), and ODMR (camera image analysis).
Over the 16 years since the introduction of the STM, no failures attributable to the STM protocol (such as experimental data corruption or deadlocks) have been reported. However, prior to the livelock-free remediation, rare and non-reproducible throughput degradation had been observed. This was a self-recovering
lock-step livelock. The phenomenon observed in the 128-thread
stress test in Section~\ref{sec:benchmarks} reproducibly triggers a persistent form of this rare legacy symptom,
and the verified true livelock-free remediation presented in this work prevents its occurrence in every verified configuration and in every
measured stress condition up to 128 threads.

\subsection{Comparison with Related Approaches}
\label{sec:comparison}

\subsubsection{Concurrent Trees and Snapshot Mechanisms}

KAME's STM shares design principles with Copy-on-Write (COW) B-trees and MVCC (Multi-Version Concurrency Control) from the database domain.

LMDB\cite{chu2011lmdb} employs a COW B+-tree with shadow paging, 
ensuring consistency through an atomic swap of the root pointer at commit time.
 Reads are lock-free, with $O(1)$ snapshot acquisition (the root pointer) and $O(\log N)$ lookups within the snapshot, where writes require path copying from the leaf to the root,
  incurring $O(\log N)$ cost. Furthermore, it imposes a single-writer constraint.

The Bw-tree\cite{levandoski2013bwtree} achieves a latch-free B-tree through delta records and CAS operations on a mapping table.
 Writes publish in $O(1)$ (a delta prepended to the mapping-table entry after the node is located), but reads require traversal of the delta chain.

Recent concurrent B+-trees designed for persistent memory\cite{yan2024concise}
also propose lock-free reads, and log-free splits in the case of CC-Tree.\cite{yan2024concise} However, all of them entail a trade-off between $O(\log N)$
write cost due to path copying or read cost due to delta chain traversal. A side-by-side comparison of
 these concurrent tree-structured data structures, including their read/write complexity and consistency model, 
 is provided in the supplementary material.

KAME's design is optimized for the access patterns of measurement software.
Each node is independently updated at high frequency, while consistent reads of subtrees are required for 
GUI display and data saving. As is conventional for B-trees, we assume an $m$-way (fanout-$m$) tree, so its height 
is $O(\log N)$ in the total number of nodes $N$. 
After the initial bundle has been
constructed, commits and snapshots perform complementary incremental
operations on affected ancestor paths. 
On the uncontended fast path, a single-node commit performs an $O(1)$ CAS in place.
If a prior snapshot had bundled the node under an ancestor, the first subsequent commit must \emph{unbundle} the affected path
(an $O(\log N)$ walk up the bundled-by chain, the same order as a COW B-tree's mandatory path copy).
Subsequent commits to the node return to the $O(1)$ fast path until a
later snapshot re-bundles that path.
Conversely, taking a consistent subtree snapshot costs $O(N)$ only for the initial cold bundle (with every node dirty).
Once warm, an intervening commit dirties just one path, so the next snapshot \emph{re-bundles} only that path at $O(\log N)$,
 and a snapshot of an unchanged subtree returns in $O(1)$. 
 Under the read-heavy access pattern of measurement software (repeated snapshots of the same subtree for GUI display and data saving), 
 the per-snapshot cost approaches $O(1)$ under stable access patterns, and trees are shallow in practice (4--7 levels). 

 By contrast, a conventional persistent COW B-tree pays its $O(\log N)$ path copy on \emph{every} single-node update, 
 which is costly when local updates are much more frequent than requests for new consistent root-tree views.
 The bundle protocol instead localizes cost to the modified path and
  materializes a reusable consistent subtree representation only when a
snapshot is requested.
Producing a consistent read-only view is not by itself difficult.
Immutable and persistently shared structures already provide such
views. The challenge addressed here is to combine on-demand subtree
snapshots with lock-free, in-place point updates and atomic
read--modify--write publication over the same subtree. To the best of
our knowledge, this particular combination has not previously been
reported.

\textbf{Perspective of dynamic structural modifications:}
In COW B-trees (LMDB), node insertion and deletion are handled as part of path copying, which is always
 consistent but incurs $O(\log N)$ cost. A single-writer constraint also applies.
The Bw-tree installs ordinary record-update deltas by applying a single
CAS to a mapping-table entry after locating the target node. Node splits,
however, use a multi-step latch-free protocol. An explicitly represented
transitional state may remain visible until the corresponding parent
update is completed, and traversing threads must recognize this state
and may help complete the structural modification.
Classical B-link trees permit searches without node locks but use locks
for updates, where an insertion may temporarily hold up to three node locks
while propagating a split. They are therefore not lock-free as complete
update data structures. 
KAME's online insertion/release 
     (Section~\ref{sec:dynamic_tree}) performs insertion and deletion via $O(1)$ CAS,
      with consistent reads recovered during bundling at $O(\log N)$ per modified path ($O(N)$ for a cold full-tree snapshot,
       approaching $O(1)$ once warm). No explicit split-in-progress detection is required,
        as the bundle protocol uniformly handles all cases including structural modifications.  

\subsubsection{Requirements Fit and Measurement-Control Frameworks}

We do not claim these choices are the only workable ones, 
but they fit the requirements of measurement software well. 
Comparable acquisition and control frameworks adopt different concurrency models and consistency scopes.
LabVIEW is built on value-oriented dataflow semantics,\cite{kodosky2020labview} and%
its queued-message architectures commonly organize mutable state within task-owning loops that exchange commands and data through queues;
the basic shared-variable abstraction does not itself provide transactions spanning multiple variables.
EPICS distributes state across a network instead of sharing an address space,\cite{dalesio1994epics}
as independently served process variables, each with its own timestamp, monitoring,
and change notification.\cite{epics_docs}
Timestamping the individual values, or the messages that carry them, makes observations comparable after the fact but does not make them atomic.
A further group of modern Python-based frameworks---qudi,\cite{binder2017qudi} the labscript suite,\cite{starkey2013labscript} and PyMoDAQ\cite{weber2021pymodaq}
---builds modular architectures on signals, messages, threads, and processes, without documenting a transactional multi-node update or a retained consistent snapshot as a native operation.
Each choice is sound for its intended use.
The present design addresses the case where parallel instrument drivers,
 shared immutable snapshots that copy no payloads, and multi-node consistency are required simultaneously.

\subsubsection{General-Purpose STM and Lifetime Management}

STM attracted considerable attention in the 2000s as a promising paradigm for concurrent programming.
The concept dates back to Shavit \& Touitou (1995),\cite{shavit1997stm} whose lock-free design required transactions and transactional storage to be statically bounded in advance;
DSTM\cite{herlihy2003dstm} added dynamically created transactions and objects under the weaker guarantee of obstruction-freedom;
Harris et al.\cite{harris2005composable} integrated composable transactions (modular blocking and choice) into Concurrent Haskell;
TL2,\cite{dice2006tl2} a widely used reference design, is lock-based, combining commit-time locking with global-version-clock validation;
and, at the language level, Clojure\cite{hickey2008clojure} offers a practical JVM STM maintaining multiple versions of managed Refs.
General-purpose STM incurs overhead from metadata management, read/write-set tracking, validation, and aborted work,
and can therefore be slower than synchronization specialized for a particular data structure or workload.

Hardware transactional memory has not provided a portable replacement: support remains
architecture-dependent, and transactions may abort because of data conflicts or buffering-capacity
limits, so a software path is necessary in any generally deployable system.\cite{yoo2013tsx}

Only a subset of STM implementations integrates transactional access with safe lifetime management for dynamically allocated objects.
In word- or object-based C/C++ STMs such as TL2, TinySTM, and NOrec, transactional locations may contain raw pointer values,
but the STM protocol alone does not guarantee the lifetime of the referenced objects;
safe reclamation requires an additional discipline such as deferred reclamation, epochs, hazard pointers, reference counting, or an STM-aware allocator.
GHC STM and Clojure Refs instead operate in garbage-collected runtimes, so transactional variables can safely retain managed references,
and a newly constructed persistent data structure can be published by atomically replacing one managed root reference---an
$O(1)$ publication step independent of the structure's size, analogous to KAME's Linkage CAS,
although the surrounding transaction and lifetime semantics differ.
Garbage-collected runtimes, however, may introduce stop-the-world or safepoint pauses whose duration depends on the collector,
heap configuration, and workload, and is not bounded tightly enough for hard real-time guarantees.
KAME provides this lifetime management in a real-time-compatible manner through lock-free split reference counting via \texttt{atomic\_shared\_ptr};
to our knowledge, no prior design simultaneously offers reference retention, optimistic $O(1)$ publication of large data,
and freedom from unbounded collector pauses. The complexity of the multi-step CAS (at least $1+2N$ per bundled commit, where $N$ is
the number of sub-Linkages: $N$ child captures during bundling, one root commit, and $N$ eventual unbundle releases, plus protocol-internal checkpoint CASes---seven or more CASes for a 3-level bundle) is justified by formal verification (Section~\ref{sec:verification}),
livelock under ultra-high contention has been eliminated through the oldest-first priority tag and per-linkage contention management,
 as demonstrated by formal and large-scale stress testing (Section~\ref{sec:verification}),
  and the production track record empirically supports the soundness of the fundamental design (Section~\ref{sec:track_record}).

\section{Conclusion}
\label{sec:conclusion}

This paper reports on
a software platform for condensed matter physics measurements,
the lock-free STM framework that forms its core,
and the results of formal verification conducted retrospectively.
The 16-year practical record rests on the STM's fit with the access patterns of measurement software:
 reads (Snapshots) are more frequent than writes, and writes to each node are typically performed by only one thread.
  Under these conditions, eliminating read-set tracking overhead and committing with a single CAS keeps the per-transaction cost low.

These schemes are summarized in the following three points:
{(i) Design and long-term operational track record of an STM based on a tree-structured object model.}
KAME's STM has been in continuous use for condensed matter physics measurements including NMR and ODMR.
Under requirements of 10~MSPS-class real-time data I/O and simultaneous control of numerous measurement instruments, 
we have demonstrated the effectiveness of the combination of a tree-structured object model and optimistic $O(1)$ snapshots.
{(ii) Design of a lock-free atomic\_shared\_ptr}.
To realize optimistic $O(1)$ snapshots, a lock-free atomic\_shared\_ptr was designed and implemented in-house in 2006. 
It employs a split reference counting scheme using tagged pointers, and provides lock-free shared pointer operations even as of 2026,
when the major standard library implementations rely on locking (an embedded lock bit with spinning or OS wait/notify).
{(iii) Formal livelock-freedom proof and exhaustive safety verification}
Exhaustive verification of protocol logic was performed using TLA+ (two layers: the lock-free shared pointer layer and the STM commit + bundle/unbundle layer),
 combined with GenMC for correctness verification under the C11/RC11 memory model.
Operations involving bundle/unbundle are ultra-multi-step CAS chains, and the state space for two or more threads is extremely large.
Accordingly, verification was restricted to bounded configurations, within which livelock-freedom of the negotiation layer was established.
Exhaustive liveness verification has been completed for the 2-thread configurations (except the 3-level dynamic release model) and, for 3 threads, the all-root contention pattern,
 at both 2- and 3-level tree depths, and is corroborated by empirical stress tests up to 128 threads with no residual livelock.
  No safety violations originating from the C++ implementation were detected in these layers.
This result is consistent with the 16-year track record of stable operation.

Real-time operation is currently under test on a real-time OS.
There, the uncontended CAS fast path is a short, bounded instruction sequence that never waits on a
preempted lock holder.
Lock-freedom alone does not bound an individual operation, but the oldest-first arbitration does:
at most $T{-}1$ older transactions can precede a given one, and each, once globally oldest, completes without
further transactional interference because its peers yield---a worst-case latency proportional to $T$ times the
bounded transaction duration, presupposing bounded transaction bodies, a bounded arbitration threshold, and
scheduling that keeps the oldest runnable.
Combined with the Python scripting interface already in production, the STM can thus serve as a shared-state
backbone for autonomous measurement systems (physical AI), in which real-time instruments and supervisory AI
agents concurrently read and write the same tree-structured state, with the acquisition path isolated from
slower supervisory computation.

\begin{acknowledgments}
The author thanks T.~Ugawa and R.~Shioya for valuable discussions.
The computation in this work has been done using the facilities of the Supercomputer Center, the Institute for Solid State Physics, the University of Tokyo (2026-A-0004).
This work was supported by JSPS KAKENHI Grant Numbers JP24K00580 and JP25K00951,
and by MEXT Supporting Pioneering Research through AI for 1,000 Discovery challenges Program (SPReAD) Japan Grant Number JPMXP1726275196.
\end{acknowledgments}

\section*{Author Declarations}
\subsection*{Conflict of Interest}
The author has no conflicts to disclose.

\section*{Data Availability}
The source code of KAME is publicly available on GitHub under the GPL v2+ license
(\url{https://github.com/northriv/KAME}). The STM core (/kamestm) and thread-local pool allocator (/kamepoolalloc) are
 dual-licensed to Apache 2.0.
The TLA+ specifications and GenMC tests used for formal verification are also included in the same repository.
The measurement data that support the findings are presented in the article and its supplementary material,
and are reproducible with the scripts, configurations, and git revisions documented therein.

\bibliography{refs}

\end{document}